\documentclass[%
  reprint,superscriptaddress,amsmath,amssymb,
  aps,nofootinbib]{revtex4-1}

\usepackage{dsfont}
\usepackage{graphicx}
\usepackage[T1]{fontenc}

\usepackage{dcolumn}
\usepackage{bm}
\usepackage{color}
\usepackage{xcolor}
\usepackage{epsfig}
\usepackage{soul}
\usepackage[colorlinks=true,urlcolor=brown,linkcolor=blue,citecolor=magenta]{hyperref}
\usepackage{natbib}
\usepackage{float}
\usepackage{footmisc}
\usepackage{siunitx}

\usepackage{subcaption}
\usepackage{caption}
\makeatletter
\g@addto@macro\normalsize{
  \setlength\abovedisplayskip{6pt plus 2pt minus 2pt}
  \setlength\belowdisplayskip{6pt plus 2pt minus 2pt}
  \setlength\abovedisplayshortskip{4pt plus 2pt minus 2pt}
  \setlength\belowdisplayshortskip{4pt plus 2pt minus 2pt}
}
\makeatother

\definecolor{lightred}{rgb}{1,0.5,0.5}
\definecolor{lightgreen}{rgb}{0.5,1,0.5}
\definecolor{lightblue}{rgb}{0.5,0.5,1}
\definecolor{lightcyan}{rgb}{0.5,0.75,0.75}
\definecolor{lightmagenta}{rgb}{0.75,0.5,0.75}
\definecolor{customgreen}{rgb}{0.494,1,0.502}

\newcommand{\htb}[1]{{\color{black} #1}}

\begin{document}

\title{Inflationary Particle Production and Implications for WIMP Substructure}

\author{María Olalla Olea-Romacho }
\email{maria_olalla.olea_romacho@kcl.ac.uk}
\affiliation{Theoretical Particle Physics and Cosmology, King’s College London, Strand, London WC2R 2LS, United Kingdom 
}

\begin{abstract}
We explore the observational consequences of resonant particle production during inflation, focusing on its impact on dark matter annihilation signals today. A transient burst of particle production generates localised features in the primordial power spectrum, enhancing the formation of compact small-scale dark matter structures known as prompt cusps. If dark matter consists of thermal WIMPs, the resulting small-scale structures substantially boost annihilation rates, leaving potentially detectable imprints in gamma-ray observations. Using 15 years of Fermi-LAT data targeting the Virgo cluster, we derive upper limits on the thermally averaged annihilation cross section $\langle \sigma v \rangle$, connecting inflationary particle production in the early universe with present-day observations constraining dark matter annihilation.
\end{abstract}

\maketitle

\section{Introduction}


Cosmic inflation~\cite{Guth:1980zm, Starobinsky:1980te} offers a powerful framework to explain the initial conditions of the universe and the origin of cosmological perturbations. In its simplest realizations, inflation generates a nearly scale-invariant primordial power spectrum of curvature perturbations. This prediction aligns well with current observations of the cosmic microwave background (CMB)~\cite{Planck:2018jri, Planck:2018nkj, ACT:2025fju, ACT:2025tim}. However, features in the power spectrum, such as bumps or oscillations, remain consistent with observational constraints at small scales and may offer a unique window into the inflaton sector and testing Planckian physics.

A compelling scenario that leads to such features arises from \emph{resonant particle production during inflation}, first introduced in Ref.~\cite{Chung:1999ve}, which can occur naturally in inflationary models derived from higher-dimensional gauge theories~\cite{Furuuchi:2015foh, Furuuchi:2020ery, Furuuchi:2020klq} or trapped inflation~\cite{Green:2009ds}. This situation requires that the inflaton couples to heavy fields that can momentarily become massless as the inflaton evolves. When this happens, the transient production of particles leads to nonlinear interactions that imprint localised features in the primordial spectrum. While it was originally proposed that this effect primarily originated from draining kinetic energy from the inflaton, suppressing $\dot{\phi}$ and enhancing perturbations via the scaling $\mathcal{P}_{\mathcal{R}}(k) \propto (H/\dot{\phi})^2$, subsequent studies have shown that the dominant contribution arises instead from nonlinear rescattering processes~\cite{Pearce:2017bdc}. In particular, the produced quanta source additional inflaton fluctuations through trilinear interactions of the form $\delta\phi\,\varphi^2$, generating bump-like features in the power spectrum~\cite{Pearce:2017bdc}.

The presence of features in the primordial power spectrum that indicate a deviation from nearly scale-invariance have attracted interest over the past decades, since they offer a means to probe high-energy physics and field interactions active during inflation~\cite{Martin:2000xs, Adams:1997de, Starobinsky:1992ts, FrancoAbellan:2023sby, Hamann:2009bz, Palma:2017wxu, Silverstein:2008sg, Flauger:2009ab, Pajer:2024ckd, Flauger:2009ab, Pajer:2024ckd, Adams:2001vc, Hazra:2014goa, Achucarro:2010da, Pi:2012gf, Braglia:2020taf, Iacconi:2023slv, Gorgulho:2025wxz}. A wide range of theoretical models predict such deviations, either in the form of broad oscillatory modulations or localised enhancements. For instance, axion monodromy models can predict oscillatory patterns across all observable scales~\cite{Silverstein:2008sg, Flauger:2009ab, Pajer:2024ckd}, while sharp features in the inflaton potential, such as steps or kinks, can induce localised oscillations or suppress power at specific scales~\cite{Starobinsky:1992ts, Adams:2001vc, Hazra:2014goa, Fairbairn:2025fko}. Multi-field inflation scenarios, where the inflationary trajectory undergoes turns or passes through phase transitions, can also generate oscillations or resonant features~\cite{Achucarro:2010da, Pi:2012gf, Braglia:2020taf, Iacconi:2023slv}.

From a phenomenological point of view, particle production during inflation can leave behind observable signatures across different experiments. Enhanced small-scale power may subsequently source second-order gravitational waves~\cite{Baumann:2007zm, Ananda:2006af, ZhengRuiFeng:2021zoz, Garcia:2025yit} or enhance the formation of small-scale dark matter structures, which  future large-scale structure (LSS) surveys such as LSST might be able to constrain~\cite{Palma:2017wxu}. Some studies have shown that single- and multi-bump scenarios in the primordial power spectrum are viable, and compatible with Planck CMB data~\cite{Naik:2022mxn}. Other analyses in a similar context~\cite{Philcox:2024jpd, Kim:2021ida} have Planck CMB data as well to search for localised hot or cold spots caused by inflationary particle production, placing upper bounds on the coupling strength of heavy fields to the inflaton and constraining particle masses up to $\mathcal{O}(100)$ times larger than the inflationary Hubble scale. Additional works have investigated whether the gravitational wave background detected by NANOGrav could be explained by resonant particle production during inflation~\cite{Gangopadhyay:2023qjr}, which induces a bump in the scalar power spectrum and subsequently leads to secondary induced gravitational waves. They show that, with suitable choices of particle mass and coupling, this mechanism can generate a spectrum consistent with the observed signal.

This paper revisits the resonant particle production mechanism as a source of localised features in the primordial power spectrum, which has important implications for dark matter substructure. The enhancement of the power at small scales can trigger the formation of prompt cusps~\cite{Diemand:2005wv,Ishiyama:2010es,Anderhalden:2013wd,Ishiyama:2014uoa,Polisensky:2015eya,Ogiya:2016hyo,Angulo:2016qof,Delos:2017thv,Delos:2018ueo,Ogiya:2017hbr,Ishiyama:2019hmh,Colombi:2020xbv, Fairbairn:2025cae}, which are dense, power-law central structures
that form during the monolithic collapse of isolated peaks
in the early dark matter density field. Galaxy clusters, being the most massive virialised structures in the Universe, serve as powerful laboratories for dark matter searches. Their deep gravitational potentials and abundant dark matter content make them promising hosts for a significant population of prompt cusps.

We aim to explore whether self-annihilating dark matter can be used as a probe of these inflationary phenomena by using 15 years of gamma-ray data from the Fermi Large Area Telescope (Fermi-LAT), covering the energy range 500 MeV to 500 GeV, to search for dark matter annihilation signals in seven nearby galaxy clusters~\cite{FermiData, Crnogorcevic:2025nwp}. 
From a theoretical perspective, weakly interacting massive particles (WIMPs) remain among the most compelling dark matter candidates, and they are a central target of experimental efforts~\cite{LZ:2024zvo}.  In general, the
inflaton decays into all available degrees of freedom including dark matter, and if the dark sector is thermally coupled to the Standard Model, it will naturally reach equilibrium and freeze out as the universe expands, yielding a relic abundance without the need for elaborate model-building. Viable WIMPs may have masses up to tens of TeV before perturbativity breaks down~\cite{Griest:1989wd}, leaving a wide and largely unexplored parameter space, in spite of the stringent limits from direct detection~\cite{LZ:2024zvo}. Moreover, and largely in agreement with these limits, models exist in which WIMPs behave as pseudo-Nambu–Goldstone bosons and their couplings to nucleons cancel at tree level, naturally suppressing direct detection signatures~\cite{Gross:2017dan}. WIMPs can annihilate into Standard Model species, so, in principle, one could trace their presence with gamma-rays. In this paper, we show that thermal WIMPs with masses up to $\mathcal{O}(1)\,\mathrm{TeV}$ and annihilation cross sections at the canonical freeze-out value $\langle \sigma v \rangle \simeq 3 \times 10^{-26}\,\mathrm{cm}^3/\mathrm{s}$ can be excluded across a broad range of inflaton–spectator couplings, approximately spanning $g \sim 0.2$–$3$, for primordial features located at scales smaller than those probed by current CMB observations. These exclusion limits arise from a substantial enhancement of the dark matter annihilation signal, driven by the formation of prompt cusps whose abundance is boosted by the amplified small-scale power spectrum generated through resonant particle production during inflation.

The remainder of this paper is organised as follows. In Section~\ref{sec:mechanism}, we review the physical mechanism of resonant particle production during inflation and its impact on the primordial power spectrum. Section~\ref{sec:prompt_cusps} describes the formation and properties of prompt cusps, compact dark matter halos seeded by enhanced small-scale fluctuations. In Section~\ref{sec:annihilation}, we compute the resulting gamma-ray signal from dark matter annihilation in galaxy clusters, incorporating the contribution of prompt cusps. Section~\ref{sec:results} presents our main results, including updated constraints on the annihilation cross section from Fermi-LAT observations. We conclude in Section~\ref{sec:conclu}.

\section{Resonant Particle Production: Physical Mechanism}
\label{sec:mechanism}
\raggedbottom   

We consider a class of inflationary models in which the inflaton field $\phi$ is coupled to a heavy spectator scalar field $\varphi$ via the interaction:
\begin{equation}
\mathcal{L}_{\text{int}} = -\frac{1}{2} g^2 (\phi - \phi_0)^2 \varphi^2.
\end{equation}
This coupling induces a time-dependent effective mass for $\varphi$, given by $m_\varphi(\phi) = g|\phi - \phi_0|$, which vanishes when the inflaton crosses the critical value $\phi = \phi_0$. At that point, the rapid variation in $m_\varphi$ results in a burst of non-adiabatic particle production, temporarily exciting the $\varphi$ field.

During inflation, the homogeneous (zero-mode) component of the inflaton field $\phi(t)$ slowly rolls down its potential, driving quasi-exponential expansion and sourcing nearly scale-invariant quantum fluctuations. However, when $\phi(t)$ approaches $\phi_0$, the $\varphi$ field becomes momentarily light, and its quanta are resonantly produced. The energy needed for this production is drawn from the inflaton's kinetic energy, leading to a transient dip in $\dot{\phi}$. The dimensionless power spectrum of curvature perturbations, $\mathcal{P}_\mathcal{R}(k)$, which quantifies the amplitude of scalar metric fluctuations as a function of comoving wavenumber $k$, scales proportionally to $\left( H / \dot{\phi} \right)^2$, where $H$ is the Hubble parameter during inflation and $\dot{\phi}$ is the time derivative of the inflaton field.

However, the direct backreaction on $\dot{\phi}$ contributes only subdominantly to the resulting feature. The dominant effect instead arises from nonlinear interactions between the produced $\varphi$ quanta and the inflaton fluctuations $\delta\phi$~\cite{Pearce:2017bdc}. In particular, the trilinear vertices, enable rescattering processes in which the $\varphi$ particles source additional inflaton perturbations. These interactions give rise to a pronounced bump in the primordial power spectrum, centered around $k \simeq 3.35,k_\star$ (i.e.~the comoving scale that crossed the horizon when the inflaton field reached the critical value $\phi = \phi_0$) followed by a damped series of oscillations. Subleading contributions from quartic terms generate smaller secondary peaks at $k \simeq 1.25 \, k_\star$~\cite{Pearce:2017bdc}.

Altogether, these contributions produce a rich structure in the power spectrum: a leading bump and a tail of oscillations superimposed on the nearly scale-invariant background. The full effect has been analytically computed using the in-in formalism in~\cite{Pearce:2017bdc}, with shape functions $f_1(x)$ and $f_2(x)$ encoding the characteristic forms of the trilinear and quartic contributions, respectively. This framework allows us to relate the imprints of resonant particle production on the matter power spectrum to the coupling $g$ and the value of $k_{\star}$.

The resulting primordial power spectrum is modeled as:
\begin{align}
\label{eq:pps}
    \mathcal{P}_\mathcal{R}(k) &= A_s \left( \frac{k}{k_*} \right)^{n_s - 1}
     \nonumber \\ &+ A_I  \left( \frac{f_1(x_\star)}{f_{1,\text{max}}} \right)
    + A_{II} \left( \frac{f_2(x_\star)}{f_{2,\text{max}}} \right),
\end{align}
where $x_\star = k/k_\star$ and the bump shape functions are:
\begin{align}
f_1(x) &= \frac{[\sin(x) - \text{Si}(x)]^2}{x^3}, \\
f_2(x) &= \frac{-2x\cos(2x) + (1 - x^2)\sin(2x)}{x^3},
\end{align}
with $\text{Si}(x)$ the sine integral. The peak locations are approximately at $k_1 \approx 3.35\,k_\star$ for $f_1$ and $k_2 \approx 1.25\,k_\star$ for $f_2$, with maximum values $f_{1,\text{max}} \approx 0.11$ and $f_{2,\text{max}} \approx 0.85$~\cite{Pearce:2017bdc}. We use the Planck 2018 best-fit values, derived from the combination of TT, TE, EE, low-$ \ell $, lensing, and BAO data, for the amplitude of scalar perturbations, $ \ln(10^{10} A_s) = 3.047 $, and the scalar spectral index, $n_s = 0.9665$, both evaluated at the pivot scale $ k_* = 0.05\,\mathrm{Mpc}^{-1}$~\cite{Planck:2018jri}.

The amplitudes of these features depend on the coupling $g$ as:
\begin{align}
A_I &\approx 6.6 \times 10^{-7} g^{7/2}, \\
A_{II} &\approx 1.1 \times 10^{-10} g^{5/2} \left[ \ln\left( \frac{g}{0.0003} \right) \right]^2,
\end{align}
valid in the perturbative regime where $g^2 \lesssim 3$~\cite{Pearce:2017bdc}.

To compute the matter power spectrum in the presence of a localised bump in the primordial power spectrum, we proceed in two steps. First, we use \texttt{COLOSSUS}~\cite{Diemer:2017bwl} to compute the matter power spectrum $P_{\Lambda\mathrm{CDM}}(k)$ for a standard $\Lambda$CDM cosmology with a nearly scale-invariant primordial spectrum. From this, we extract the transfer function by defining:
\begin{equation}
T(k) \equiv \sqrt{\frac{P_{\Lambda\mathrm{CDM}}(k)}{\mathcal{P}_{\mathcal{R},\,\Lambda\mathrm{CDM}}(k)}},
\end{equation}
where $\mathcal{P}_{\mathcal{R},\,\Lambda\mathrm{CDM}}(k) = A_s \left( \frac{k}{k_\ast} \right)^{n_s - 1}$ is the smooth primordial curvature power spectrum with the values of the amplitude and spectral index indicated above.

We then use the modified primordial power spectrum $\mathcal{P}_\mathcal{R}(k)$ (see Eq.~\ref{eq:pps}) including the features from resonant particle production, and compute the corresponding matter power spectrum via
\begin{equation}
P(k) = \mathcal{P}_\mathcal{R}(k)\, T^2(k).
\end{equation}
The matter power spectrum $P(k)$ is the key quantity from which we infer the statistical properties of prompt cusps and their impact on dark matter annihilation, which we will describe in the next section.

\section{Prompt cusp properties}
\label{sec:prompt_cusps}
Prompt cusps are highly concentrated dark matter structures that emerge from the direct gravitational collapse of isolated overdensity peaks in the early Universe. In contrast to Navarro-Frenk-White (NFW) halos, which form through hierarchical merging and exhibit a characteristic central slope $\rho \propto r^{-1}$, prompt cusps develop much steeper inner density profiles scaling as $\rho \propto r^{-3/2}$~\cite{Delos:2019mxl,Delos:2022bhp}. This profile is established at the time of collapse and encodes information about the height and curvature of the primordial peak from which the cusp originates.

Not every overdensity peak evolves into a prompt cusp. Peaks that are rapidly accreted into larger structures may be disrupted by mergers, while others fail to collapse before becoming incorporated into a more massive halo. Numerical simulations suggest that small halos embedded within larger hosts often retain their inner structure, but prompt cusps that experience major mergers are typically erased via dynamical friction~\cite{Delos:2022yhn}. Taking these effects into account, we follow~\cite{Delos:2022yhn} and assume a survival probability of $f_{\rm surv} \approx 0.5$.

In standard $\Lambda$CDM cosmologies, the formation of the smallest halos is bounded from below by the free-streaming scale, $\lambda_{\rm fs}$, which encodes the minimal scale at which dark matter fluctuations survive the residual thermal motion. Perturbations on smaller scales are erased, leading to a sharp cutoff in the power spectrum at $k_{\rm fs} \sim \lambda_{\rm fs}^{-1}$. Halos forming near this threshold are not originated via mergers of smaller halos but collapse directly from primordial peaks.

To model this suppression, we include a damping factor of the form $\exp(-k^2/k_{\rm fs}^2)$ in the matter power spectrum. The corresponding cutoff scale depends on the dark matter mass $m_\chi$ and the kinetic decoupling temperature $T_{\rm kd}$~\cite{Green:2005fa}:
\begin{align}
k_{\mathrm{fs}} &\approx 1.70 \times 10^6\, \mathrm{Mpc}^{-1} 
\left( \frac{m_{\chi}}{100\,\mathrm{GeV}} \right)^{1/2} \\
&\times \left( \frac{T_{\mathrm{kd}}}{30\,\mathrm{MeV}} \right)^{1/2}
\left[ 1 + \frac{\ln(T_{\mathrm{kd}} / 30\,\mathrm{MeV})}{19.2} \right]^{-1}.
\end{align}
Throughout this work, we fix the kinetic decoupling temperature to $T_{\rm kd}=1\,\mathrm{GeV}$ as a benchmark value; the results are largely insensitive to this choice because the dependence of Eq.~\eqref{eq:Jcusp} on $T_{\rm kd}$ is only logarithmic (see below).
In scenarios where inflationary particle production amplifies small-scale power, halo formation is no longer governed by the free-streaming limit. Instead, halos tend to form first at roughly the wavenumber corresponding to the bump in the power spectrum. These enhanced modes become nonlinear earlier than those near $k_{\rm fs}$ and collapse into bound structures first. It is therefore the scale of the inflationary feature, not the free-streaming limit, that sets the stage for early halo formation.

These cosmologies feature an enhanced abundance of high-amplitude peaks, which increases the probability of prompt cusp formation~\cite{Ralegankar:2023pyx}. The cusps are described by a density profile of the form
\begin{equation}
\rho(r) = A\, r^{-3/2},
\end{equation}
bounded from below by a core radius $r_{\rm core}$ and truncated from above at a radius $r_{\rm cusp}$. The former arises from microphysical limits, while the latter reflects the initial size of the perturbation.

Since the annihilation rate in such steep profiles diverges in the $r \to 0$ limit, the inner region must be regularised. The core radius $r_{\rm core}$ is determined by the phase-space density constraint set by the dark matter's thermal velocity at kinetic decoupling. Analytic estimates~\cite{Delos:2022yhn}, supported by numerical simulations~\cite{Maccio:2012qf}, give
\begin{equation}
r_{\mathrm{core}} \simeq 0.34\, G^{-2/3} \left( \frac{m_{\chi}}{T_{\mathrm{kd}}} \right)^{-2/3} \bar{\rho}(a_{\mathrm{kd}})^{-4/9} A^{-2/9},
\end{equation}
where $\bar{\rho}(a_{\mathrm{kd}})$ denotes the mean background density at kinetic decoupling. 

The normalization constant $A$ is set by the scale and time of collapse and can be inferred statistically from the power spectrum~\cite{Delos:2019mxl}:
\begin{equation}
A \simeq 24\, \bar{\rho}(a_{\mathrm{coll}})\, (a_{\mathrm{coll}} R)^{3/2},
\end{equation}
where $a_{\mathrm{coll}}$ is the scale factor at collapse, estimated via the ellipsoidal collapse criterion~\cite{Sheth:1999su}, and $R$ is the characteristic comoving radius of the peak. This is defined as
\begin{equation}
R = \left| \frac{\delta}{\nabla^2 \delta} \right|^{1/2},
\end{equation}
where both the density contrast $\delta$ and its Laplacian are evaluated at the location of the peak.

The outer radius of the cusp is determined by the scale of the initial perturbation and is given by~\cite{Delos:2019mxl}:
\begin{equation}
r_{\mathrm{cusp}} \simeq 0.11\, a_{\mathrm{coll}} R.
\end{equation}

Prompt cusps are thus expected to be among the densest dark matter structures in cold dark matter cosmologies. Simulations indicate that a subset can survive disruption and remain intact until today~\cite{Delos:2022bhp}.

To simulate the statistical distribution of prompt cusps seeded by a given matter power spectrum, we used the public code developed in~\cite{Delos:2019mxl,Delos:2022yhn}. This tool combines BBKS peak statistics~\cite{Bardeen:1985tr}, and uses the ellipsoidal collapse criterion~\cite{Sheth:1999su} to determine the fate of peaks and return the corresponding asymptotic profiles for those that successfully collapse.

\htb{We note that when the primordial power spectrum has a pronounced bump with residual small-scale power, BBKS peak statistics tend to associate the characteristic peak scale with the smallest available scale in the spectrum, typically the free-streaming length. Numerical simulations~\cite{Delos:2017thv, Delos:2018ueo} instead indicate that, in such scenarios, prompt cusps form at the scale of the bump. For small or moderate bumps, however, the annihilation signal remains sensitive to residual small-scale power, making the inclusion of a free-streaming cutoff more appropriate. Since the transition between these regimes is not sharply defined, we consider two approaches: one in which we include the exponential suppression at the free-streaming scale, appropriate when cusp formation is still influenced by small-scale modes, and one in which the power spectrum is truncated at the bump scale, forcing cusp formation to occur at the bump, and closer to what is observed in simulations. We treat the difference as an uncertainty associated with the BBKS approach.}

\section{Annihilation signal}
\label{sec:annihilation}
The steep central profiles of prompt cusps significantly amplify the dark matter annihilation rate relative to standard halo models. Although cusps make up only a small fraction of the total dark matter mass (about $5\%$), their high central densities yield a significant contribution to the annihilation signal~\cite{Delos:2022bhp, Fairbairn:2025cae}.

To quantify their contribution, we adopt the formalism introduced in~\cite{Delos:2022bhp} and extended in~\cite{Crnogorcevic:2025nwp}. The differential flux of photons from dark matter annihilation is given by
\begin{equation}
    \frac{\rm{d}^2\Phi}{\rm{d} \Omega\, dE} = \frac{\langle \sigma v \rangle}{8\pi m_\chi^2} \frac{\rm{d} N_\gamma}{ \rm{d} E} \frac{\rm{d} \rm J}{\rm{d}\Omega},
\end{equation}
where \( \langle \sigma v \rangle \) denotes the velocity-averaged annihilation cross section, \( \rm{d} N_\gamma / \rm{d} E \) the photon spectrum per annihilation, and \( \rm{dJ}/\rm{d}\Omega \) the astrophysical \( \rm{J} \)-factor. In the presence of prompt cusps, the total \( \rm{J} \)-factor acquires an additional contribution,
\begin{equation}
    \frac{\rm{dJ}}{\rm{d}\Omega} = \int d\ell\, \rho(r) \left[ \rho(r) + f_{\rm surv} f_{\rm tidal}(r) \rho_{\rm eff, 0} \right],
\end{equation}
where the integration is carried out along the line of sight \( \ell \). The first term accounts for the smooth dark matter distribution, for which we adopt a NFW profile. The second term captures the annihilation signal from surviving prompt cusps embedded in the halo. Here, \( \rho_{\rm eff, 0} \) is the effective squared density contributed by the initial population of cusps. We neglect the radial suppression factor \( f_{\rm tidal}(r) \), which accounts for tidal stripping of cusps near the halo center. As shown in~\cite{Crnogorcevic:2025nwp}, its contribution to the total annihilation signal remains at the percent level and does not significantly impact the inferred constraints.

The quantity \( \rho_{\rm eff, 0} \) is computed statistically by sampling density peaks from the enhanced power spectrum:
\begin{equation}\label{eq:rhoeff0}
     \rho_{\rm eff, 0}
    \equiv \frac{\int_\mathrm{cusps}\rho^2 \rm d V}{M}
    =n_\mathrm{peaks}\langle j\rangle/\bar\rho_0,
\end{equation}
where \( n_{\rm peaks} \) is the comoving number density of collapsed peaks~\cite{Delos:2019mxl}, and \( \bar\rho_0 \simeq 33\, M_\odot\,\mathrm{kpc}^{-3} \). We generate $N = 10^4$ peaks and compute the average annihilation integral per cusp,
\begin{align}\label{eq:Jcusp}
    j \equiv \int_\mathrm{cusp}\rho^2 \rm d V = 4\pi A^2\left[\frac{1}{3}+\rm {ln}\left(\frac{r_{\rm cusp}}{ r_{\rm core}}\right)\right],
\end{align}
with $A$ the amplitude of the density profile and $r_{\rm core}$, $r_{\rm cusp}$ the inner and outer boundaries as defined in the previous section.

For our analysis, we focus on the Virgo cluster, which yields the strongest constraints among the nearby clusters studied in~\cite{Crnogorcevic:2025nwp}. We model Virgo with an NFW profile with scale radius \( r_s = 335.10 \, \mathrm{kpc} \) and scale density \( \rho_s = 305646\, M_\odot\,\mathrm{kpc}^{-3} \), located at a distance \( d_L = 15.46 \, \mathrm{Mpc} \), and subtending an angle \( \theta_{200} = 6.32^\circ \) at the virial radius~\cite{DiMauro:2023qat}.

Using our estimate of \( \rho_{\rm eff, 0} \), we compute the \( \rm{J} \)-factors and rescale the exclusion limits on the annihilation cross section from Ref.~\cite{Crnogorcevic:2025nwp} as a function of dark matter mass.

\section{Results}
\label{sec:results}
We consider three benchmark scenarios at small scales, beyond the sensitivity of current CMB measurements, each defined by a distinct combination of the coupling constant $g$ and the feature scale $k_\star$:
\begin{align}
g &= 0.2, & k_\star &= 100\,\mathrm{Mpc}^{-1}, \nonumber\\
g &= 0.4, & k_\star &= 300\,\mathrm{Mpc}^{-1}, \nonumber\\
g &= 0.8, & k_\star &= 1000\,\mathrm{Mpc}^{-1}. 
\end{align}
These scales are far below the resolution of current CMB observations and therefore remain unconstrained by CMB and LSS measurements. The chosen benchmarks lie within the perturbative regime and span a range from moderate to stronger enhancements of the primordial power spectrum, driven by increasingly larger couplings to the inflaton. Each case corresponds to a localised feature appearing at progressively smaller physical scales.
\begin{figure}[H]
    \centering
    \includegraphics[width=1.00\linewidth]{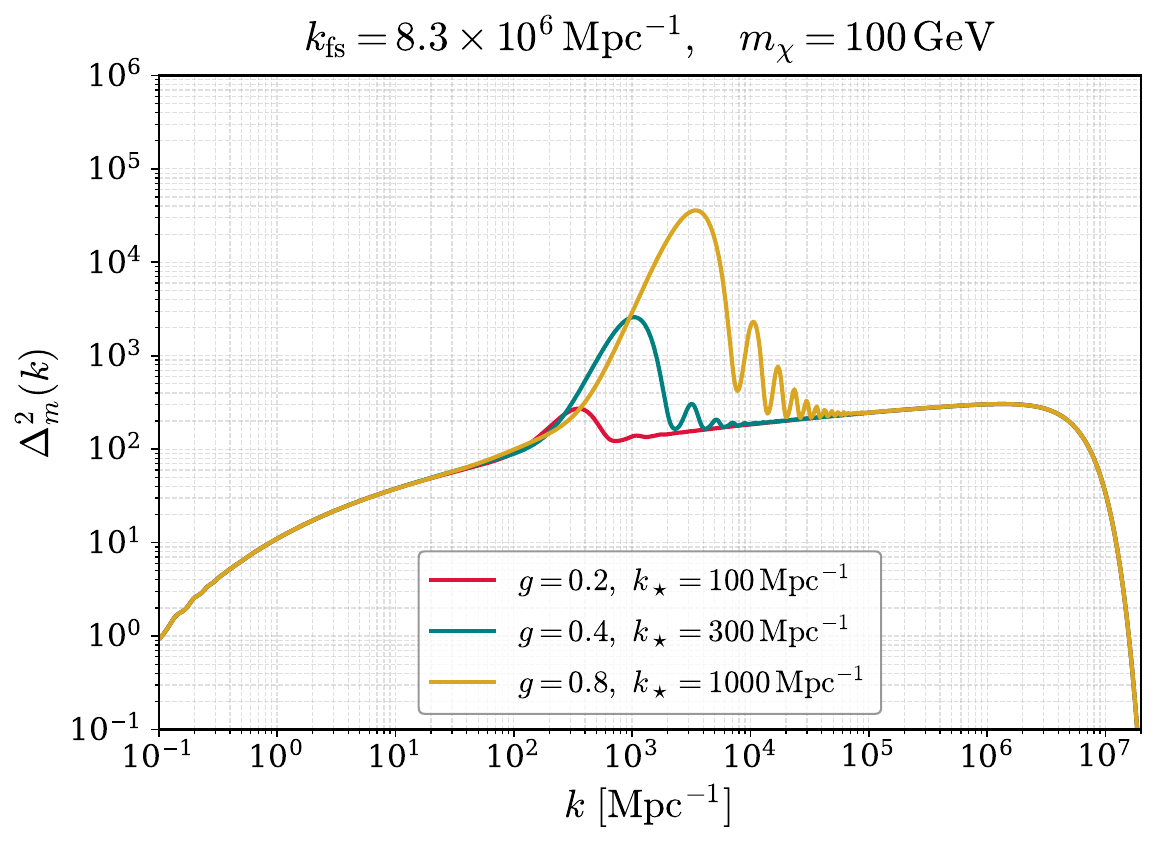}
    \caption{
Dimensionless matter power spectrum $ \Delta^2_m(k) $ as a function of comoving wavenumber $ k $, for three benchmark scenarios of resonant particle production during inflation. Each curve corresponds to a different coupling $ g $ and characteristic scale $ k_\star $, illustrating how stronger couplings lead to larger enhancements and sharper features. The free-streaming cutoff for a dark matter particle with mass $ m_\chi = 100\,\mathrm{GeV} $ and $T_{\rm kd}= 1 \, \rm{GeV}$ is shown at $ k_{\mathrm{fs}} = 8.3 \times 10^6\,\mathrm{Mpc}^{-1} $, suppressing power at the smallest scales.
}
\label{powerspectra}
\end{figure}

In Fig.~\ref{powerspectra}, the dimensionless matter power spectrum exhibits three distinct features: a suppression at large wavenumbers due to the free-streaming of dark matter, a localised bump, and a series of damped oscillations. The exponential cutoff at $ k \gtrsim k_{\rm fs} $ reflects the erasure of small-scale perturbations by thermal motion, which prevents structure formation below the free-streaming length. The main bump in the power spectrum at $ k \sim k_\star $ arises predominantly from nonlinear rescattering: quanta of a spectator field $ \varphi $, produced non-adiabatically during inflation, source inflaton fluctuations through the trilinear interaction $ \delta\phi\,\varphi^2 $~\cite{Pearce:2017bdc}. This results in a localised enhancement in the curvature power spectrum. The damped oscillations at $ k > k_\star $ arise also mainly from nonlinear rescattering processes. These interactions lead to characteristic oscillatory integrals that imprint a decaying pattern in the power spectrum beyond the main bump.
The amplitude increases with the coupling strength $ g $, as stronger couplings lead to more efficient particle production and a larger rescattering contribution to the inflaton perturbations.

\begin{figure*}[t] 
    \centering
    \begin{subfigure}[t]{0.322\textwidth}
        \includegraphics[width=\textwidth]{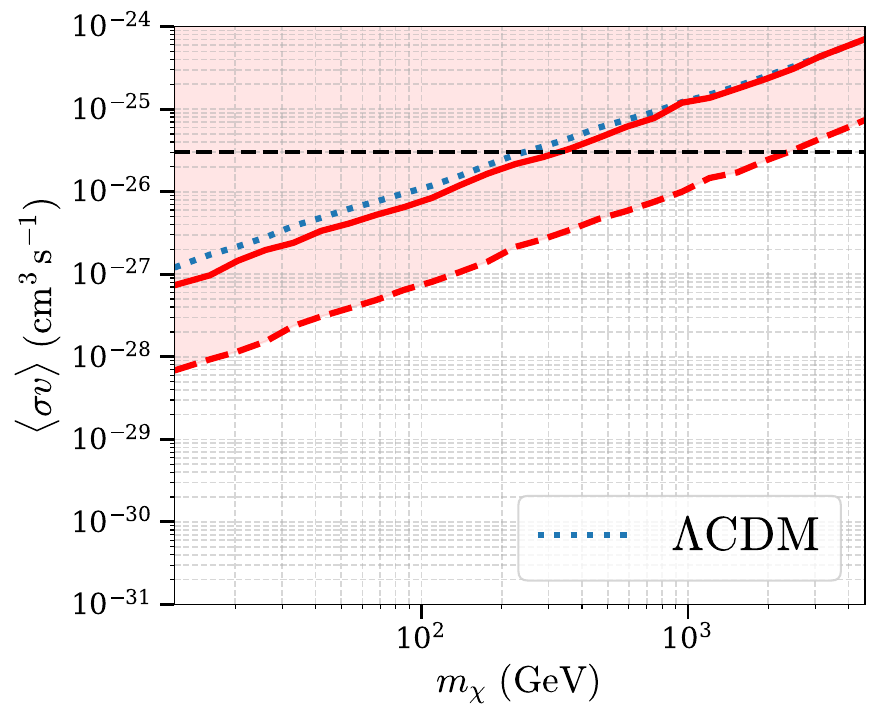}
        \caption{\small $g = 0.2$, $k_\star = 100\,\mathrm{Mpc}^{-1}$}
        \label{fig:virgo_g02}
    \end{subfigure}
    \hfill
    \begin{subfigure}[t]{0.32\textwidth}
        \includegraphics[width=\textwidth]{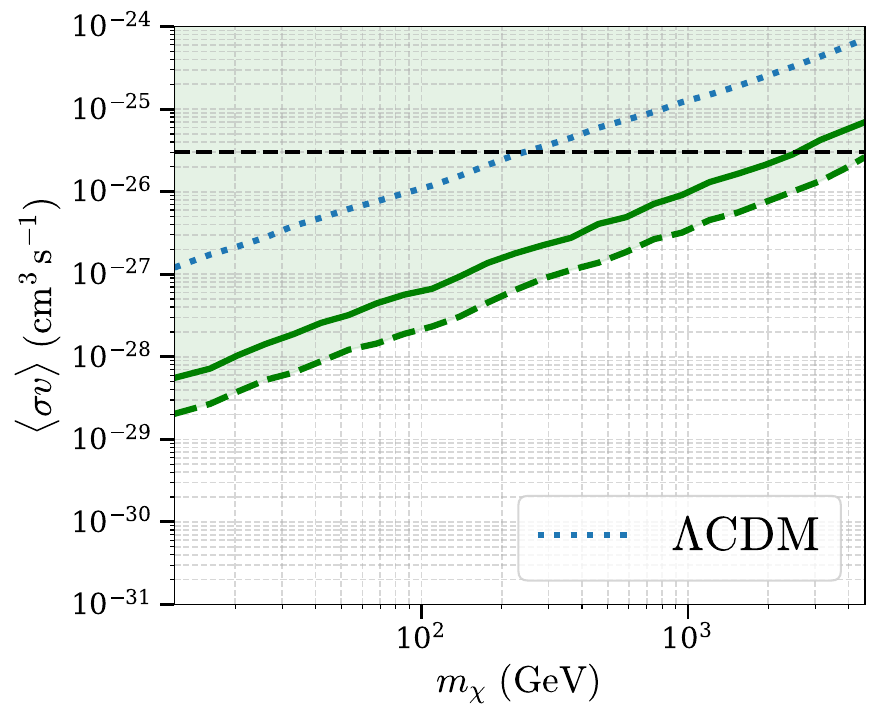}
        \caption{\small $g = 0.4$, $k_\star = 300\,\mathrm{Mpc}^{-1}$}
        \label{fig:virgo_g04}
    \end{subfigure}
    \hfill
    \begin{subfigure}[t]{0.32\textwidth}
        \includegraphics[width=\textwidth]{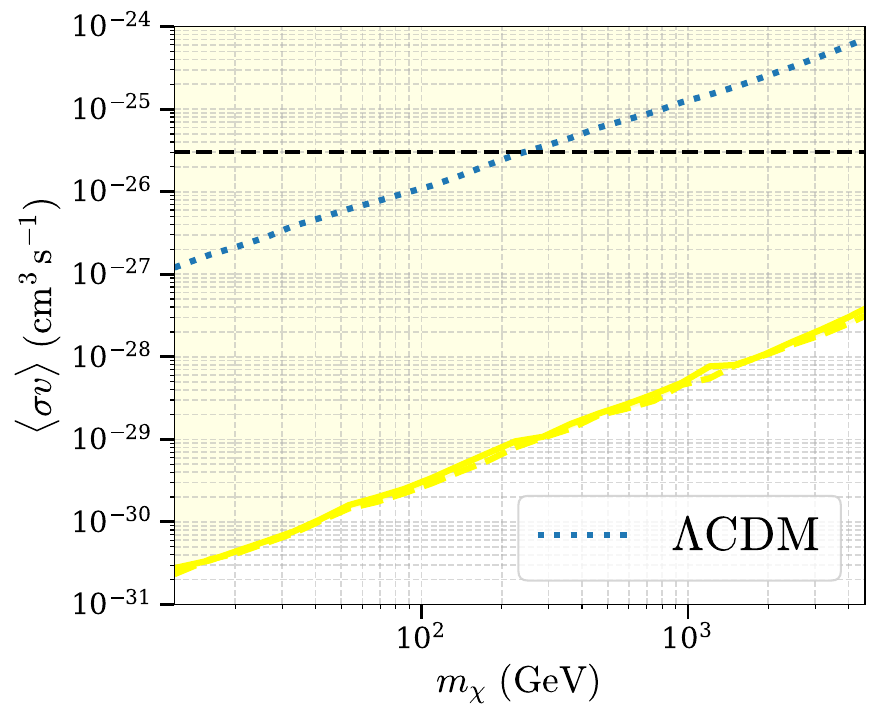}
        \caption{\small $g = 0.8$, $k_\star = 1000\,\mathrm{Mpc}^{-1}$}
        \label{fig:virgo_g08}
    \end{subfigure}
    \vspace{-1mm}
    \caption{\small Upper limits on the dark matter annihilation cross section $\langle \sigma v \rangle$ into $b\bar{b}$ from the Virgo cluster for three benchmark scenarios of resonant particle production during inflation. \htb{The blue dotted curve shows the standard $\Lambda$CDM prediction, while the other two curves correspond to a primordial power spectrum with a bump. The solid lines assume prompt cusp formation at the bump scale (spectrum cut after the bump), whereas the dashed lines include the free-streaming cutoff. The horizontal dashed black line indicates the thermal relic cross section.}}
    \label{fig:virgo_triptych}
\end{figure*}

Fig.~\ref{fig:virgo_triptych} presents the upper bounds on the thermally averaged annihilation cross section $\langle \sigma v \rangle$ derived from Fermi-LAT gamma-ray data of the Virgo cluster, assuming 100$\%$ annihilation into $b\bar{b}$, for the three (small scale) benchmark scenarios of resonant particle production shown in Fig.~\ref{powerspectra}. We chose the $b\bar{b}$ channel since this is the most constraining one according to Ref.~\cite{Crnogorcevic:2025nwp}. Similar analyses could be done for other channels such as $\tau \tau$ or $W^{-}W^{+}$, yielding weaker constraints. The $ b\bar{b} $ annihilation channel provides the most stringent constraints on dark matter in this analysis due to its production of a high multiplicity of gamma rays with a broad, soft energy spectrum. This spectrum aligns well with the peak sensitivity range of the Fermi-LAT telescope. In contrast, channels like $ \tau^+ \tau^- $ and $ W^+ W^- $ produce fewer photons or have higher energy thresholds, leading to weaker constraints~\cite{Ando:2012vu}. The $ W^+W^- $ channel is also kinematically suppressed below the electroweak scale, further limiting its constraining reach at low masses.  Furthermore, galaxy clusters such as Virgo are especially effective targets for probing dark matter annihilation in the presence of enhanced small-scale structure. Unlike dwarf spheroidal galaxies, clusters harbor significantly more dark matter and a richer population of subhalos, which amplifies the cumulative contribution from prompt cusps. This increased signal, combined with the sensitivity of Fermi-LAT to the broad gamma-ray spectrum produced by $b\bar{b}$ annihilation, results in stronger constraints than those obtained from dwarf spheroidals or the isotropic gamma-ray background~\cite{Crnogorcevic:2025nwp}.

Each scenario is characterised by different values of the coupling $g$ and characteristic scale $k_\star$, which determine the amplitude and location of the bump in the primordial power spectrum, and consequently the abundance of prompt cusps. \htb{For each benchmark, we show results obtained both by retaining the exponential suppression at the free-streaming scale (dashed) and by truncating the spectrum at the bump scale (solid). For smaller bumps, the difference between the two cases is more pronounced, because the power integrated at scales above the bump remains comparable to that contained in the feature itself, so the annihilation signal is still sensitive to how small-scale modes are treated. As the bump becomes more prominent, the integrated power is increasingly dominated by the feature, and the annihilation rate becomes insensitive to whether the spectrum is truncated at the bump or at the free-streaming scale. In all cases, the resulting uncertainty remains limited to at most an $\mathcal{O}(10)$ effect on the annihilation limits.}
 Across all panels, we observe an improvement in sensitivity relative to the standard $\Lambda$CDM case (dashed lines) once inflationary particle production is included (solid lines). These constraints exclude the thermal relic cross section $\langle \sigma v \rangle \approx 3 \times 10^{-26}\,\text{cm}^3\text{s}^{-1}$, for thermal WIMPs with masses $\gtrsim 3 \, \rm{TeV}$ for the three benchmark scenarios explored. In this set-up, indirect detection experiments are sensitive not only to dark matter properties but also to the physics of the inflationary epoch.
The physical interpretation of this result lies in the enhanced formation of compact minihalos at the scale of the feature. \htb{Larger values of the coupling $g$ amplify the power spectrum bump, causing overdensities to collapse earlier and resulting in denser prompt cusps.} Meanwhile, the position of the bump, set by $k_\star$, controls the typical comoving size of the overdensities and influences their collapse time. While smaller-scale features at higher $k$ correspond to structures that enter the horizon earlier, their eventual collapse depends on whether the amplitude is sufficient for these scales to become nonlinear first. As a result, the annihilation flux is shaped by the interplay between how many cusps form and how concentrated they are.

Importantly, these results establish a direct observational link between inflationary dynamics and indirect detection. If a gamma-ray excess consistent with such cuspy structures were observed in the future, it could be interpreted as evidence for features in the primordial power spectrum originating from non-trivial inflationary interactions. Conversely, the absence of a signal places meaningful limits on the allowed coupling strength $g$ and the dark matter mass $m_{\chi}$, given this combined scenario, connecting early universe physics with late-time astrophysical observations.

\section{Conclusions}
\label{sec:conclu}

In this work, we explored the observational consequences of resonant particle production during inflation and its implications for dark matter annihilation signals. A transient coupling between the inflaton and heavy spectator fields generates localised features in the primordial power spectrum, enhancing the abundance of small-scale density peaks that can collapse into compact dark matter substructures known as prompt cusps.

We analysed how these inflationary features affect the formation and survival of prompt cusps and how their presence boosts the annihilation signal from self-annihilating dark matter. For thermal WIMPs with canonical annihilation cross sections, the amplified small-scale power spectrum leads to an overproduction of dense substructures, resulting in gamma-ray fluxes that would exceed current Fermi-LAT limits for galaxy clusters. In particular, we find that thermal WIMPs with masses beyond $\mathcal{O}(1) \, \mathrm{TeV}$ and annihilation cross sections at the canonical freeze-out value $\langle \sigma v \rangle \simeq 3 \times 10^{-26},\mathrm{cm}^3 \,\mathrm{s}^{-1}$ can be excluded across a broad range of inflaton–spectator couplings, approximately spanning $g \sim 0.2$–$3$, for primordial features located at scales smaller than those probed by current CMB observations. 

This study demonstrates a direct connection between early-universe inflationary dynamics and present-day indirect dark matter searches. By linking the microphysics of inflationary particle production to gamma-ray observations, we show that the absence of dark matter annihilation signals can already place meaningful limits on the coupling between the inflaton and heavy spectator fields. Future gamma-ray measurements will further refine these constraints, providing a new observational handle on inflationary scenarios with non-trivial particle production.

\section*{Acknowledgments}
I am grateful to Thomas Biekötter for useful comments on the manuscript. I am supported by the STFC under grant ST/X000753/1. I also want to acknowledge support of the grant PID2024-161668NB-I00.

\bibliography{references}
\bibliographystyle{apsrev4-1}

\end{document}